\documentclass[abstract=true]{scrartcl}

\usepackage{graphicx}
\usepackage{subfigure}
\usepackage{amssymb}
\usepackage{amsmath}

\usepackage[bookmarks,backref=true,linkcolor=black]{hyperref} 
\hypersetup{
  pdfauthor = {},
  pdftitle = {},
  pdfsubject = {},
  pdfkeywords = {},
  colorlinks=true,
  linkcolor= black,
  citecolor= black,
  pageanchor=true,
  urlcolor = black,
  plainpages = false,
  linktocpage
}

\title{Mapping Tasks to Interactions for Graph Exploration and Graph Editing on Interactive Surfaces}

\author{S. Gladisch, U. Kister, C. Tominski, R. Dachselt, H. Schumann}

\begin{document}
\maketitle

\begin{abstract}Graph exploration and editing are still mostly considered independently and systems to work with are not designed for todays interactive surfaces like smartphones, tablets or tabletops. 
When developing a system for those modern devices that supports both graph exploration and graph editing, it is necessary to 1) identify what basic tasks need to be supported, 2) what interactions can be used, and 3) how to map these tasks and interactions.
This technical report provides a list of basic interaction tasks for graph exploration and editing as a result of an extensive system review. 
Moreover, different interaction modalities  of interactive surfaces are reviewed according to their interaction vocabulary and further degrees of freedom that can be used to make interactions distinguishable are discussed.
Beyond the scope of graph exploration and editing, we provide an approach for finding and evaluating a mapping from tasks to interactions, that is generally applicable.
Thus, this work acts as a guideline for developing a system for graph exploration and editing that is specifically designed for interactive surfaces.
\end{abstract} 

\section{Introduction}
\label{introduction}


Graphs play an important role in many application domains. 
Common examples are the analysis of social networks or the illustration of biological dependencies.
When working with abstract data like graphs, visualization systems are often used.
There are numerous systems with a broad repertoire of functions supporting users in exploring graphs.
These systems have in common, that they are designed for desktop environments where users interact with mouse and keyboard only.
Although modern interactive surfaces (e.g., smartphones, tablets, tabletops) have big potential for visualization and replace desktop computers more and more, they are currently not supported by graph exploration systems.
Beside graph exploration, there are various visualization systems supporting users in editing graphs.
Although most of these systems are also designed for desktop environments, there are first approaches that utilize interactive surfaces \cite{Arvo2005, Arvo2006, Chen2003, Frisch2009, Frisch2010, Hammond2006, Plimmer2010}.

The visual analysis of graphs includes tasks for exploring and editing the data \cite{Kandel2011}.
Visualization systems supporting both aspects to the same extent are currently not provided.
Users often have to work with several systems side-by-side and switch between them frequently, which complicates analysis.
Thus, there is a need for visualization systems that support both graph exploration and graph editing.
This technical report serves as a guideline to conceive such a system and focuses on answering the following questions:

\begin{enumerate}
\item	\textbf{What tasks to support? }The functionality needed by such a system can be determined through the tasks users have to accomplish when exploring or editing graphs.
There are taxonomies that classify tasks involved in graph analysis \cite{Kerren2013, Lee2006} and user intents in general \cite{Yi2007}, but it is not clear what set of tasks should be generally provided to support exploring and editing of graphs. 
Thus, a semantic description of \textit{basic tasks} for graph exploration and graph editing is needed.
As a result of extensive reviews of different existing systems and task taxonomies, this report provides a list and classification of basic tasks. 

\item	\textbf{What interactions can be used? } Modern human-computer interaction with interactive surfaces encompasses different interaction modalities that are inspired from real world interaction. 
These include touch-, pen- and tangible interaction.
Especially for graph exploration and editing, there is no systematic description of what interactions are generally possible with either of these modalities. 
This report reviews interaction vocabularies for the mentioned modalities and further discusses degrees of freedom for making individual interaction gestures distinguishable.

\item	\textbf{How to map tasks to interactions?} In order to conceive interactions, the tasks need to be mapped to gestures. 
Beside a generic mapping approach that is applicably in general, this report states the difficulty of such a mapping and provides examples.
\end{enumerate}

Before going into detail concerning the three question, we will introduce the required basics.

\section{Terms and Systems}
\label{basics}
In this section, we clarify terms used in the following and describe visualization systems we reviewed for our approach.

\subsection*{Terms}
\paragraph{Graph} A graph is an abstract data structure and can be defined as ordered pair $G = (V, E)$ comprising a set $V$ of nodes together with a set $E$ of edges, where each edge is related to two nodes.
\paragraph{Interaction functionality} With the term interaction functionality, we refer to the range of operations or functions that can be performed interactively via the user interface of a system. A specific operation for instance can be the selection of a node.
\paragraph{Interaction modality} An interaction modality is a communication channel between the human and the computer, through which users can provide input. An example is the mouse or keyboard.
\paragraph{Gesture} With the term gesture, we refer to a concrete user action in a specific modality, for instance a left click on a mouse button.

\subsection*{Systems for graph exploration and editing}
To identify a set of basic tasks, it was necessary to analyze existing systems for graph exploration and graph editing.
The systems we considered are briefly described in the following with their individual emphasis.

\begin{figure*}[t!]
\centering
\subfigure[Graph exploration using CGV]{
\label{subfig:cgv}  
\includegraphics[height=0.135\textheight]{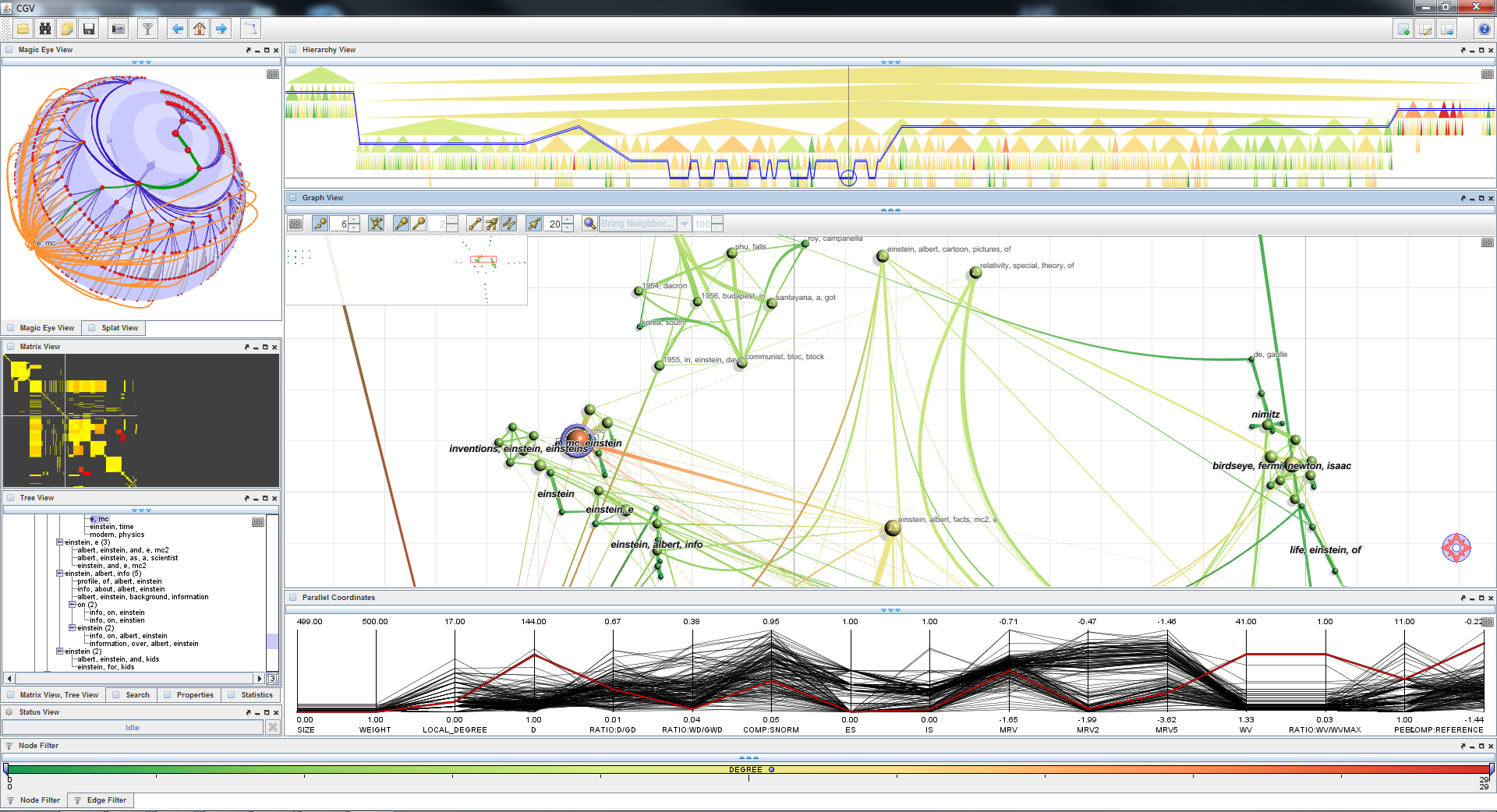}}
\subfigure[Graph exploration using Gephi]{
\label{subfig:gephi}
\includegraphics[height=0.135\textheight]{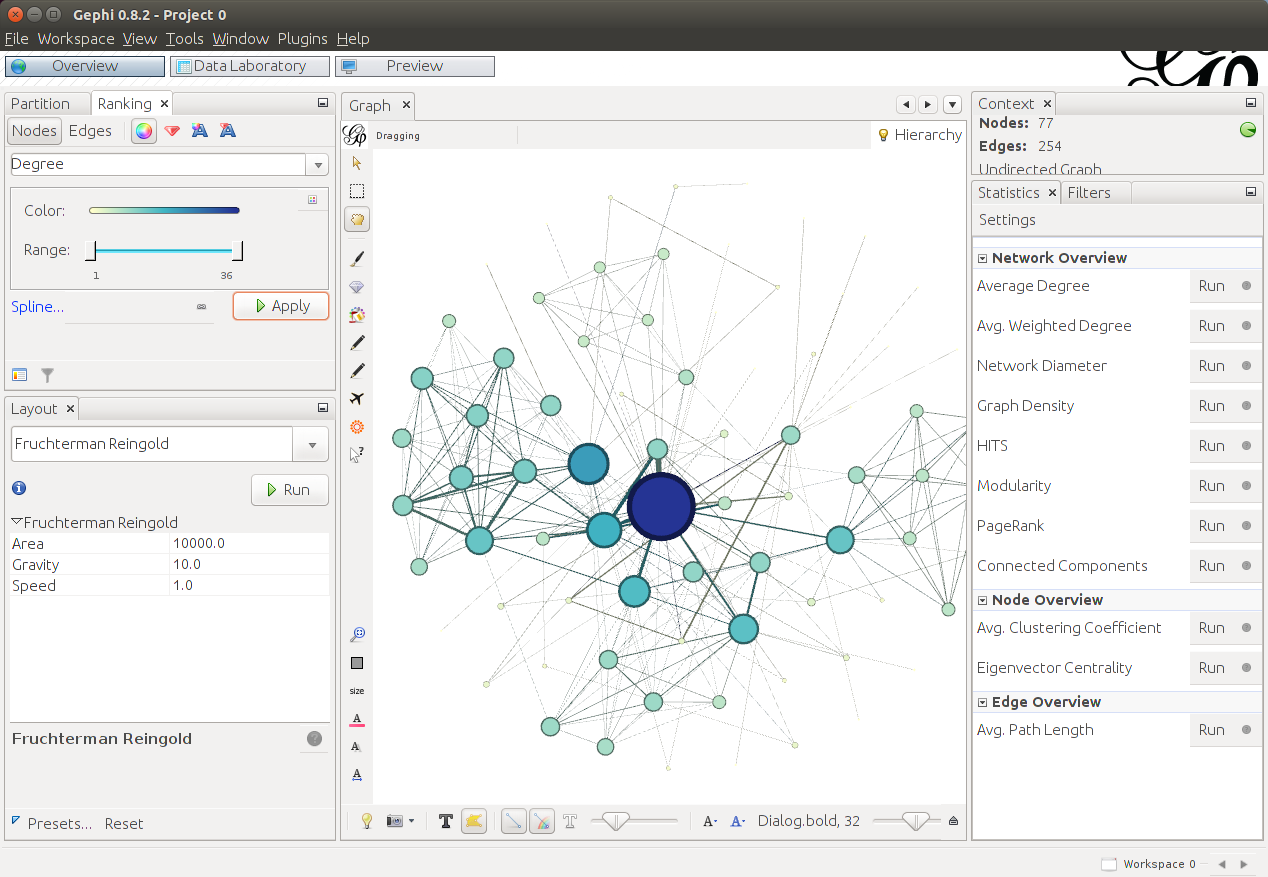}}
\subfigure[Graph exploration using Tulip. Detail from figure 16 in \cite{Auber2004}]{
\label{subfig:tulip}
\includegraphics[height=0.135\textheight]{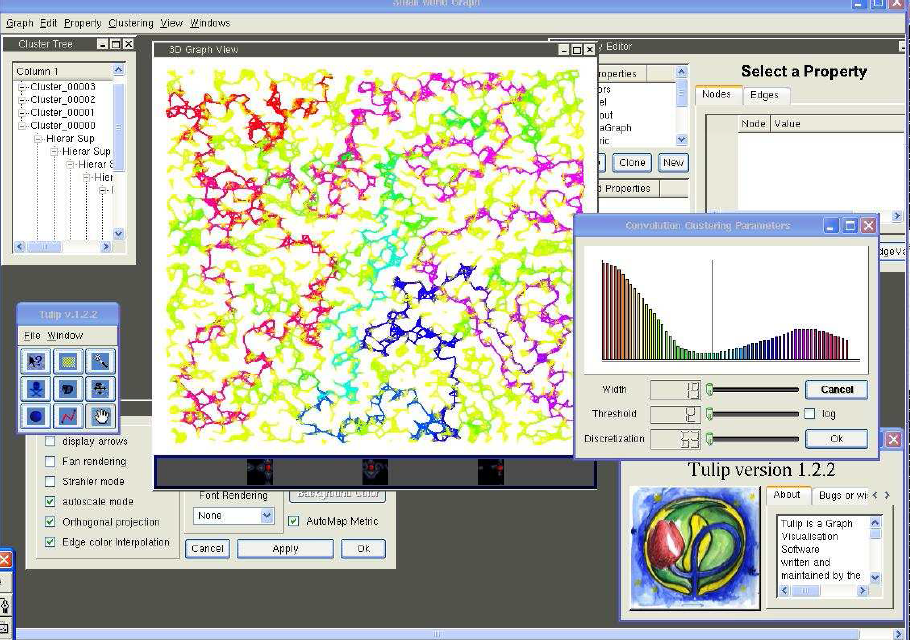}}
\subfigure[Graph exploration using Cytoscape]{
\label{subfig:cytoscape}
\includegraphics[height=0.12\textheight]{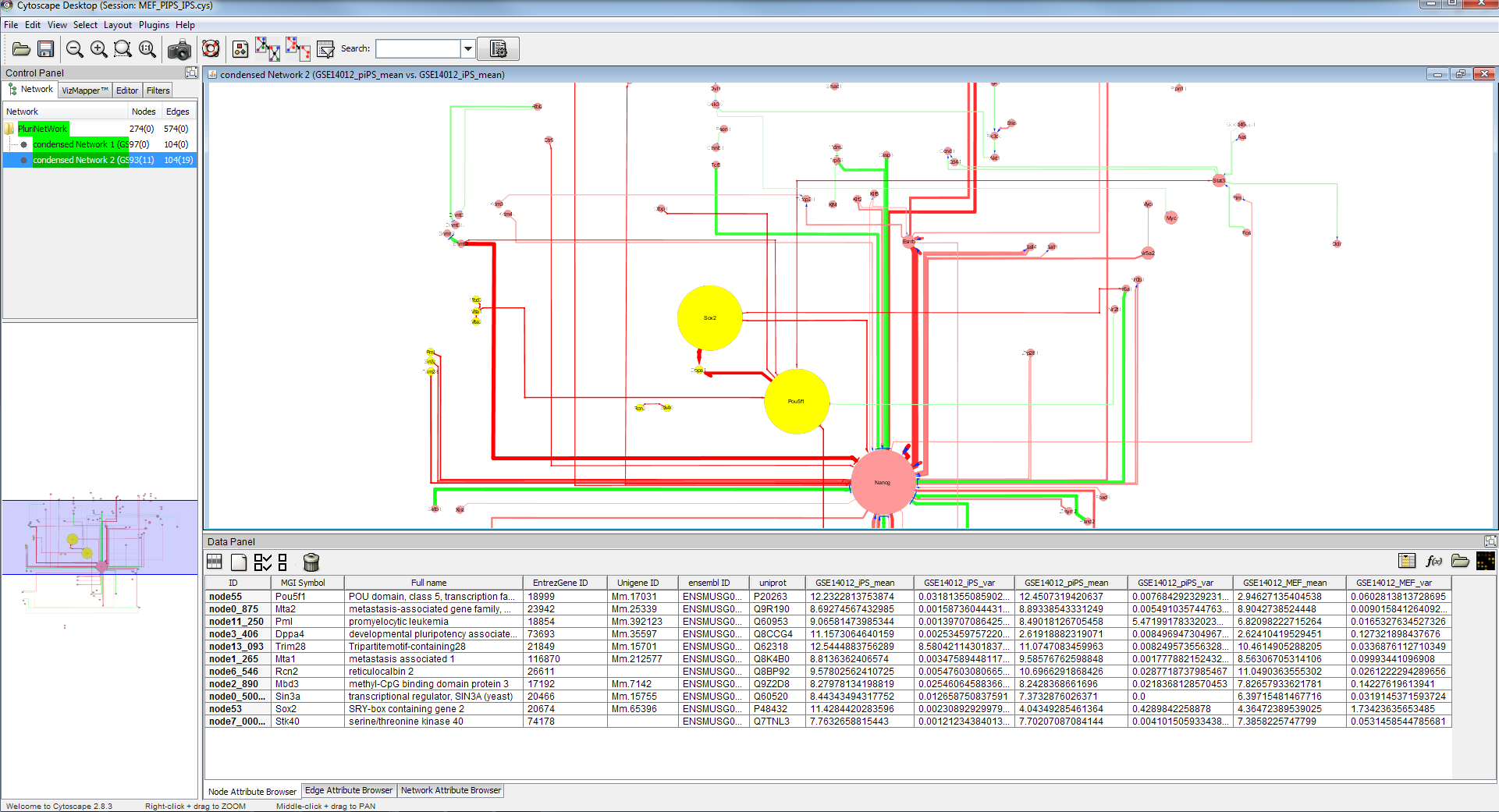}}
\subfigure[Graph exploration using Pajek]{
\label{subfig:pajek}
\includegraphics[height=0.12\textheight]{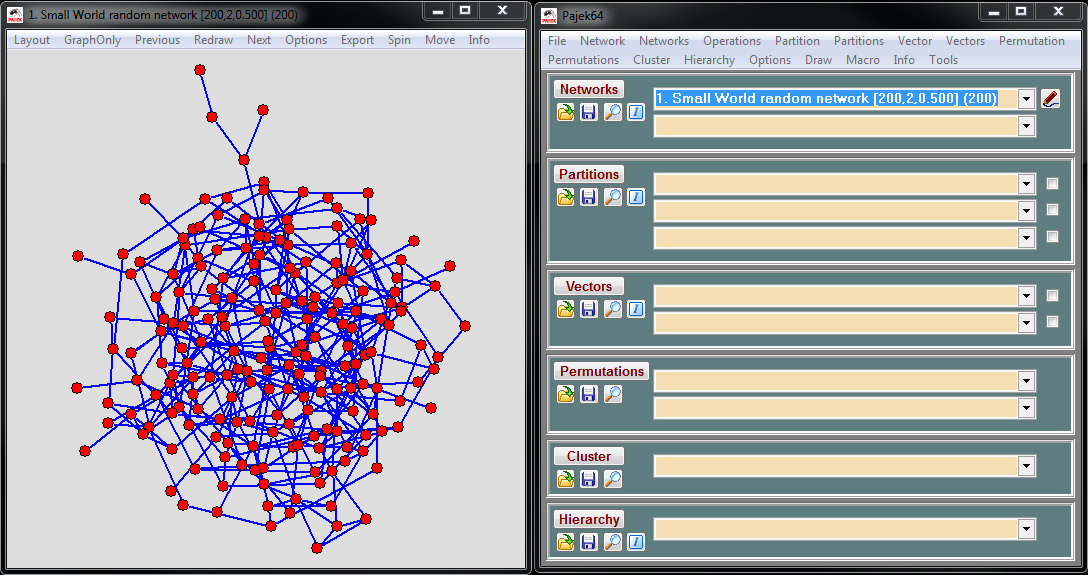}}
\subfigure[Graph exploration using Nodes3D]{
\label{subfig:nodes3D}
\includegraphics[height=0.12\textheight]{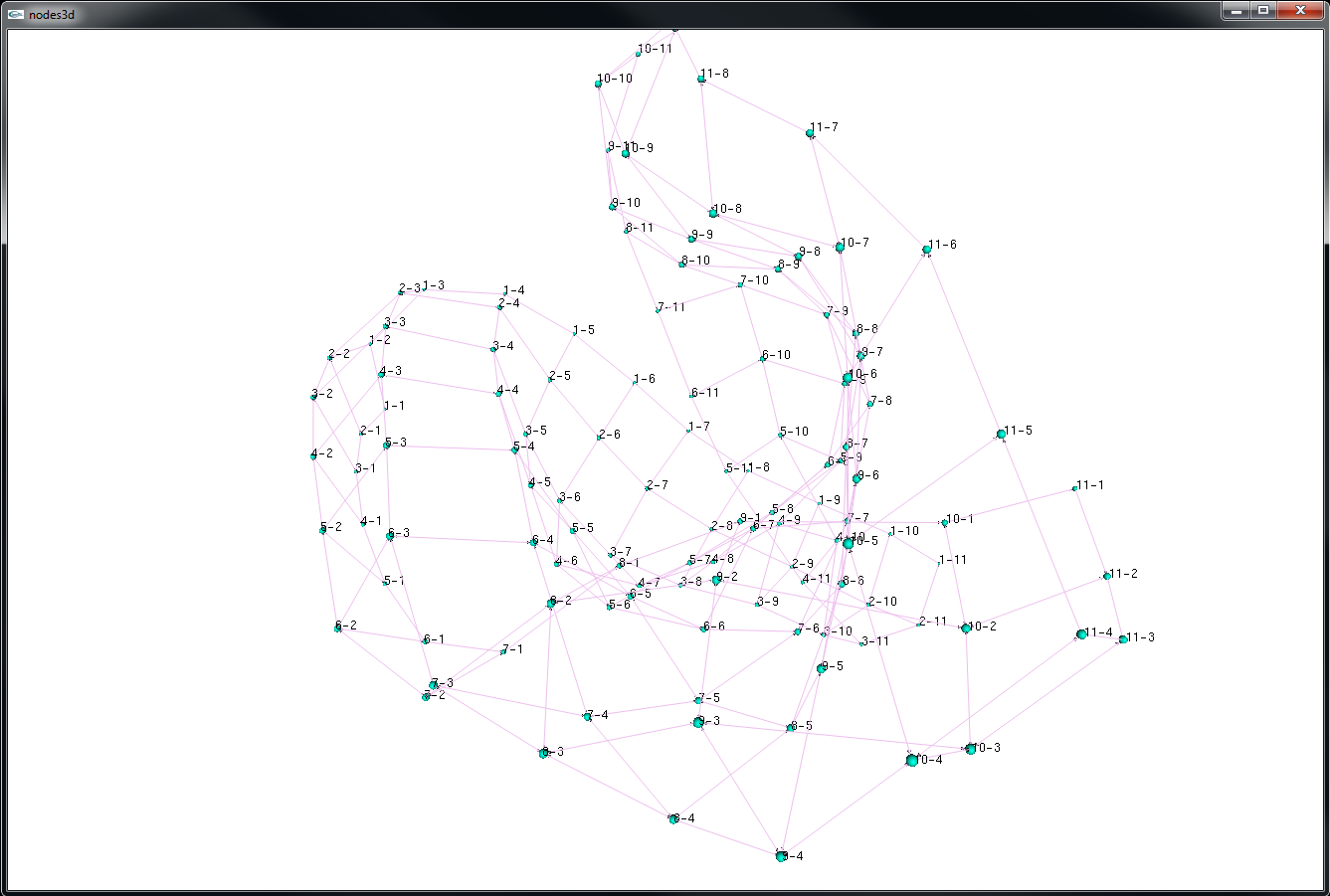}}
\caption{
Different graph visualization systems that support graph exploration: (a) CGV, (b) Gephi, (c) Tulip, (d) Cytoscape, (e) Pajek (f) Nodes3D.}
\label{fig:systemsvis}
\end{figure*}

\subsubsection*{Visualization Systems for Graph Exploration}
Several system exist that enable visual graph exploration. Some of these systems originated from research (e.g., CGV \cite{Tominski2009}, Gephi \cite{Mathieu2009}, Tulip \cite{Auber2004} and Cytoscape \cite{Shannon2003}). All of these are designed as desktop applications whereby users interact with mouse and keyboard through a WIMP (windows, icons, menus, pointers) interface. For our approach, we analyzed a set of well-known systems.

\textbf{CGV} is a system with emphasis on interaction, that uses multiple interlinked views \cite{Tominski2009}. This system is able to represent clustered graphs and allows exploration of local regions of interest using interactive lens techniques. 
\textbf{Gephi} is a graph exploration framework with broad functionality \cite{Mathieu2009}. It supports users in analyzing general graphs, clustered graphs, graphs where attributes are associated with nodes or edges, and dynamic graphs.
A general system and library for visual graph exploration with the focus on research prototyping as well as development of end-user applications is \textbf{Tulip} \cite{Auber2004}. Similar to CGV and Gephi, Tulip is able to display different aspects of the data at the same time using multiple interlinked views.

A graph visualization system especially designed for the visualization and analysis of molecular interaction networks and biological pathways is \textbf{Cytoscape} \cite{Shannon2003}. Using cytoscape it is possible to integrate networks with annotations, gene expression profiles and other state data.
\textbf{Pajek} is a graph visualization system with the focus on supporting graph abstraction by recursive decomposition of large graphs into several smaller graphs, and providing a selection of graph drawing algorithms \cite{Batagel03}.
With the graph visualization system \textbf{Nodes3D}, it is possible to visualize graphs with 3D node-link diagrams. The main focus of this system is the analysis of brain data.
Figure \ref{fig:systemsvis} presents screenshots of the mentioned systems displaying sample data sets.

\subsubsection*{Visualization Systems for Graph Editing}
Direct editing of node-link diagrams is supported by a variety of systems. Most of these systems are not limited to graphs and node-link diagrams but support diagramming in general. Anyhow, many different types of diagrams are based on objects that are connected with lines, for instance UML-Diagrams, business process diagrams, or entity-relationship diagrams, and can be interpreted as node-link diagrams. Similar to existing graph visualization systems, these systems are designed for desktop environments. User interaction is mainly based on drag and drop gestures using mouse input. We selected six established systems for analysis.

\textbf{yEd Graph Editor} is a system with focus on node-link diagram editing. Besides manual editing it provides a library of good layout algorithms. 
Similar to yEd Graph Editor, \textbf{GoVisual Diagram Editor}'s focus is on node-link diagrams and the support of automatic layouting. It supports clustered graphs via nested node-link diagrams.
\textbf{MS Visio} and \textbf{Dia Diagram Editor} are diagramming applications that provide templates for different kinds of diagrams, e.g., flow diagrams, organization charts or block diagrams. Their focus is on supporting users in designing diagrams quickly and with little effort.
\textbf{Visual Paradigm} and \textbf{Enterprise Architect} are two systems that support visual modelling. Although their focus is on modeling with UML diagrams, they also support other technologies like business process modelling with BPMN.

\begin{figure*}[t!]
\centering
\subfigure[Graph editing using yEd Graph Editor]{
\label{subfig:yed}  
\includegraphics[height=0.125\textheight]{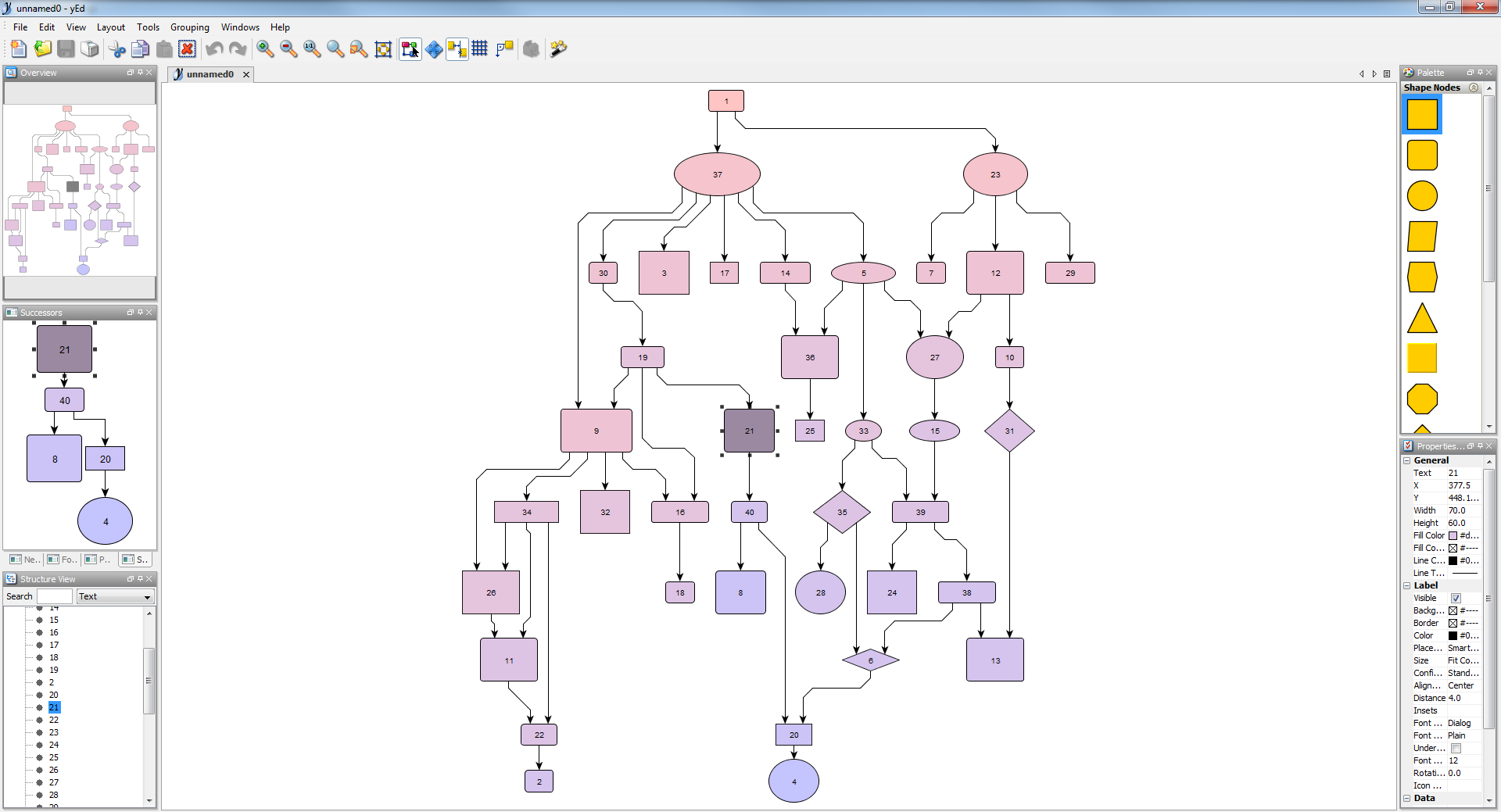}}
\subfigure[Graph editing using MS Visio]{
\label{subfig:visio}
\includegraphics[height=0.125\textheight]{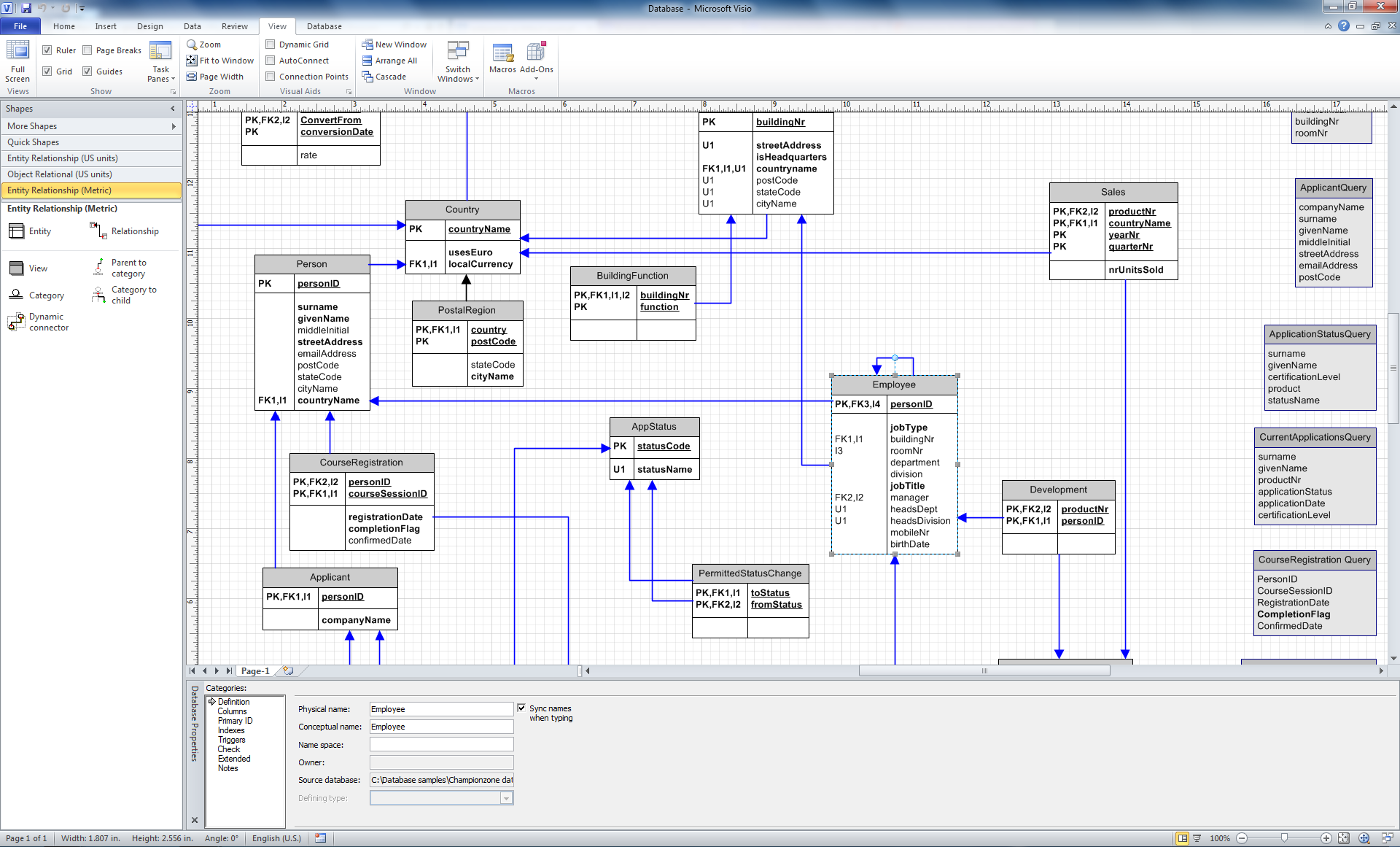}}
\subfigure[Graph editing using GoVisual Diagram Editor]{
\label{subfig:gde}
\includegraphics[height=0.125\textheight]{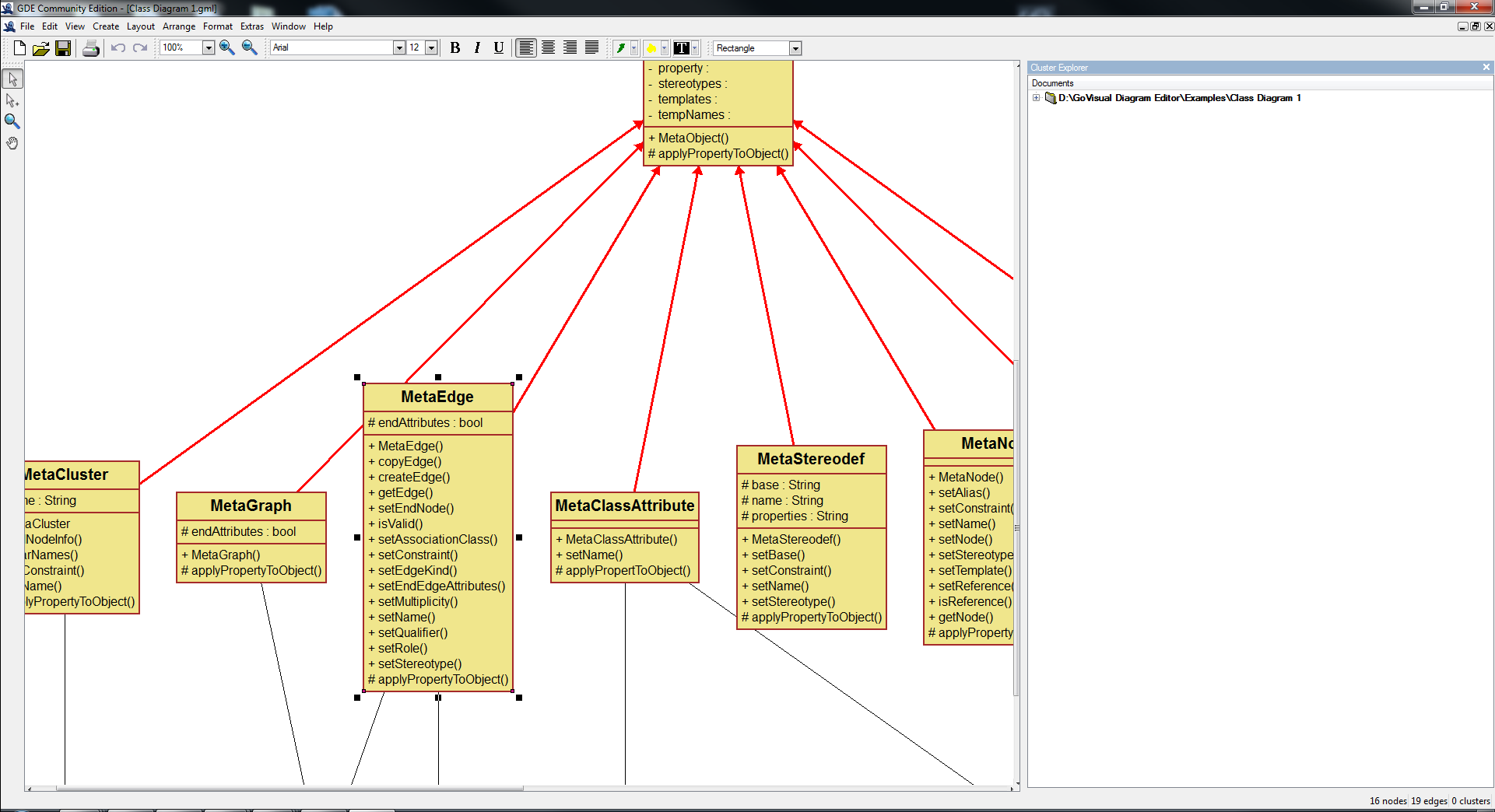}}
\subfigure[Graph editing using Dia Diagram Editor]{
\label{subfig:dia}
\includegraphics[height=0.12\textheight]{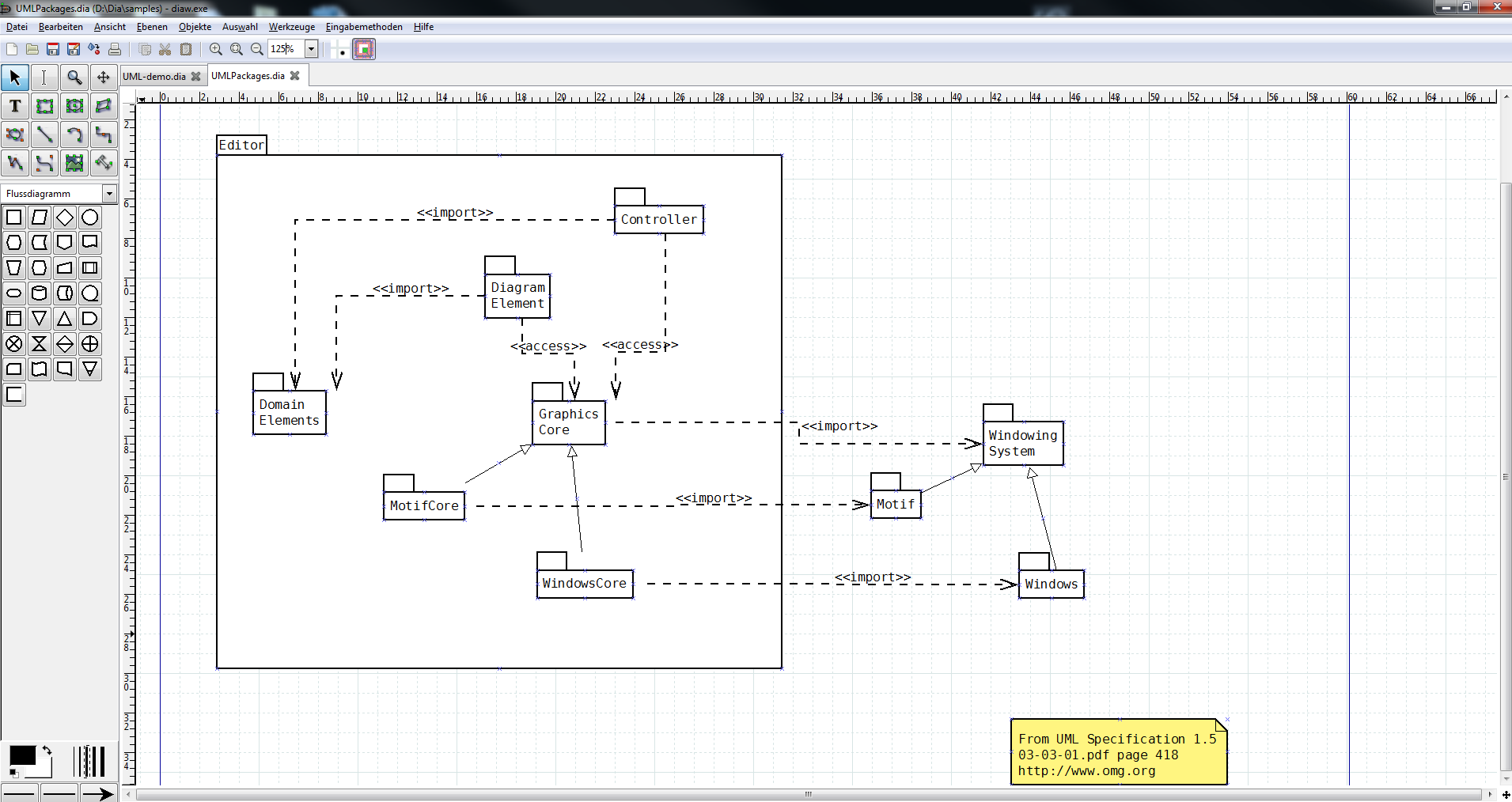}}
\subfigure[Graph editing using Visual Paradigm]{
\label{subfig:vp}
\includegraphics[height=0.12\textheight]{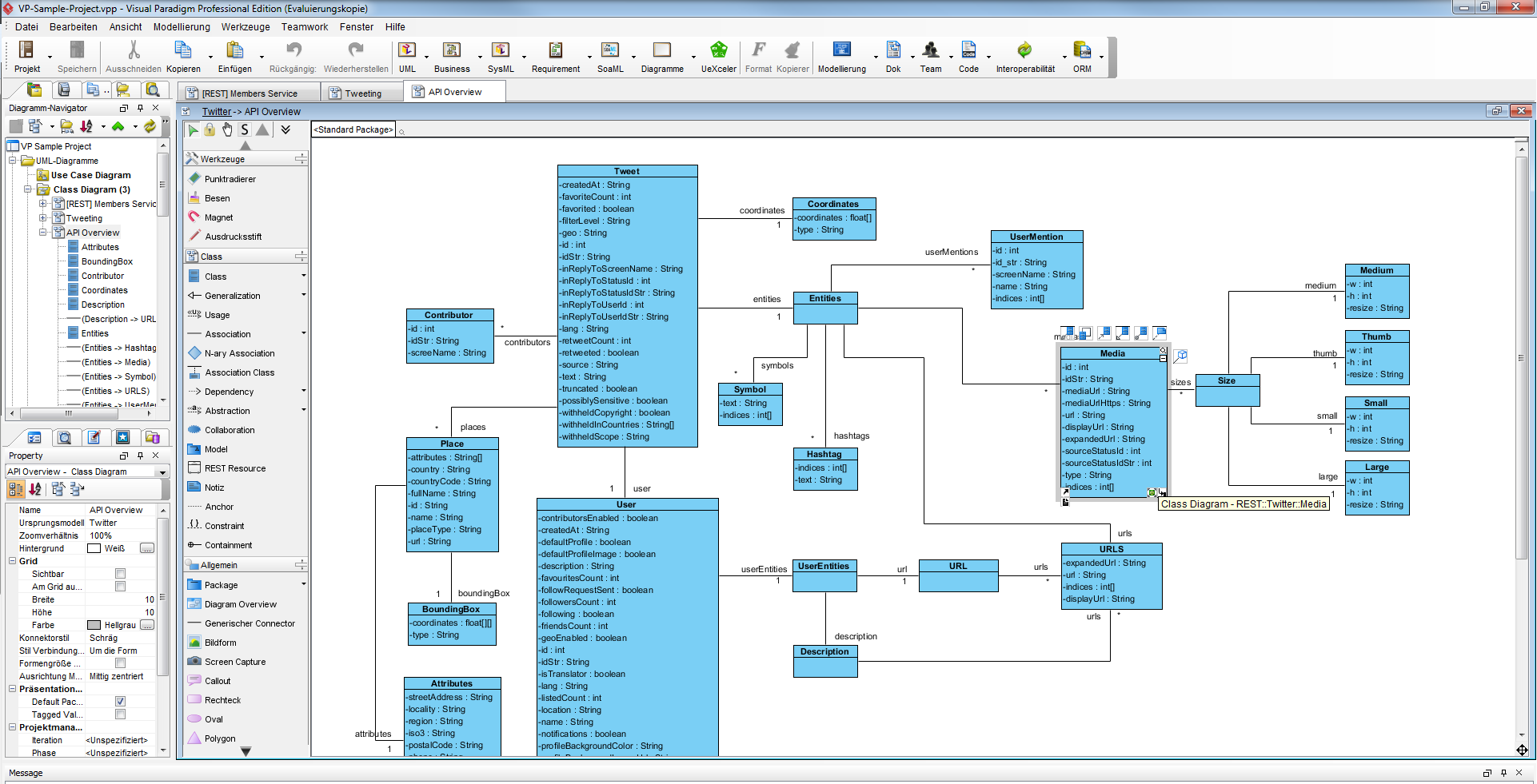}}
\subfigure[Graph editing using Enterprise Architect]{
\label{subfig:ea}
\includegraphics[height=0.12\textheight]{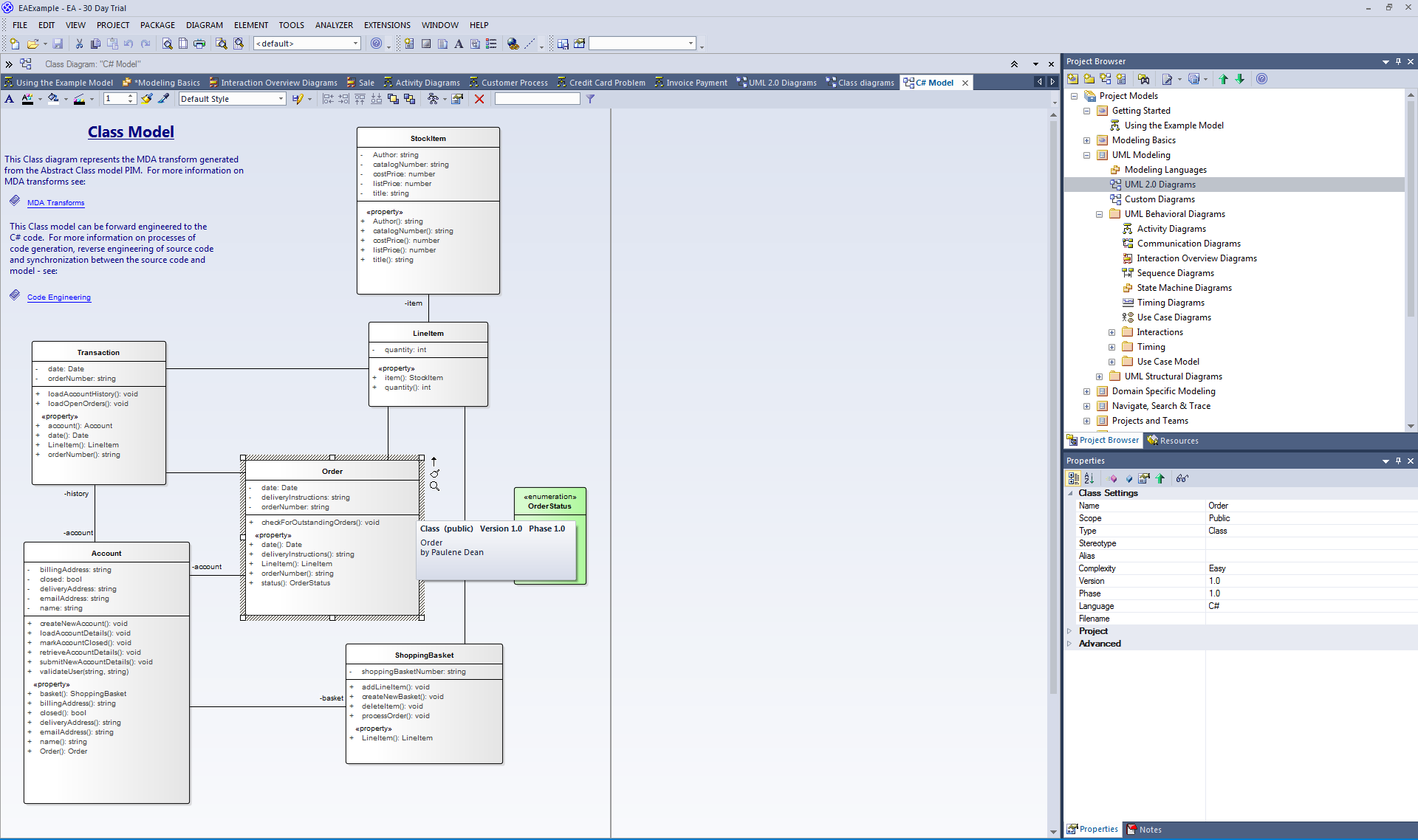}}
\caption{
Different visualization systems that support graph editing: (a) yEd Graph Editor, (b) MS Visio, (c) GoVisual Diagram Editor, (d) Dia Diagram Editor, (e) Visual Paradigm, (f) Enterprise Architect.}
\label{fig:systemsed}
\end{figure*}

Figure \ref{fig:systemsed} presents screenshots of the described system displaying node-link diagrams of a sample data set.
In the following section, we identify basic tasks for graph exploration and editing based on analyzing the systems listed above.
\section{Basic Tasks for Exploring and Editing Graphs}
\label{tasks}

In order to identify a set of basic tasks, we analyzed the graph visualization systems listed in section \ref{basics} according to their functionality.
Tasks that could be interactively accomplished were documented and categorized.
With regard to graph exploration, we used the well-established categories from Yi et al.~\cite{Yi2007} which are: select, explore, reconfigure, encode, abstract and elaborate, filter, and connect.
With regard to graph editing such a well-established categorization does not exist. Hence, we distinguish between the following classes for editing: create, insert, update, delete, explore, select, and miscellaneous categories.
In order to separate basic tasks from tasks for specific scenarios, we used the following pragmatic approach:
We consider a task to be basic, if at least three systems share functions to accomplish this task.
Since the focus of the reviewed systems differs and thus also their functionality, this approach can reveal tasks that are supported across systems, i.e., basic tasks.
We further categorized user interactions to operate the tasks according to their interaction modes \cite{Spence2007} in stepped, continuous or composite interaction.
The category stepped interaction describes interactions consisting of a series of individual steps (e.g., a mouse double click), continuous interactions are operated without interruptions (e.g., a mouse drag).
Composite interactions are a combination of stepped and continuous interactions (e.g., a mouse click followed by a mouse drag).

\subsection*{Basic Tasks for Graph Exploration}
In the following, a categorization of basic tasks along with the most frequently used interaction mode (in brackets) for every task is presented. 
Individual tasks are described and examples for interaction techniques supporting these tasks in the reviewed systems are given.

\paragraph{Select Tasks:} By performing selection tasks, users mark data items of interest to keep track of them.
When working with graphs, items to be selected can be nodes, edges, subgraphs, associated data attributes or any set of these.
\begin{itemize}
\item	Select node (mostly stepped) / Deselect node (always stepped):  When selecting a node, users specify their interest in this node. Selecting a node is an essential task and it contributes to various higher level tasks, e.g. to view a nodes attributes or characteristics. With deselecting a node, the users signalizes that this node is not of interest anymore. In most of the systems, a node can be selected with a mouse click or a selection frame.
\item	Select multiple nodes (varying between all three) / Deselect multiple node (always stepped): These task are accomplished similarly to the above tasks just for sets of nodes.
\item Temporary select node (varying between all three) / Temporary select edge (varying between all three). With these tasks individual nodes or edges are specified as temporary interesting. An exemplary interaction technique supporting this task is to hover a node or an edge in the visualization with a mouse cursor. The temporary interest is often supported by highlighting the affected node or edge.
\end{itemize}

\paragraph{Explore Tasks:} By performing exploration tasks, users examine global characteristics and different subsets of the data set (details), for instance a subgraph of a node-link diagram that is currently placed off-screen.
\begin{itemize}
\item Pan view (mostly continuous): This navigation task enables users to move the current view parallel to the view plane in order to examine different subsets of the visualized data. Dragging scrollbars or moving the background of the visualization itself are commonly used interaction techniques in this regard.
\item Center view (always stepped): A task to center the view on a specific point of interest. This can be the center of the initial view or the position of a node for example. This task is mostly supported by selecting a menu option or clicking on a dedicated button.
\item Rotate view (mostly composite): A task to rotate the view around a specific axis with a certain angle. In some systems, this can be achieved trough a menu selection followed by a mouse drag that specifies the rotation angle.
\item Zoom view (mostly continuous): This navigation task enables users to zoom into or out of specific content in the current view. Rotating the mouse wheel is a common way to achieve zooming.
\end{itemize}

\paragraph{Reconfigure Tasks:} When reconfigure tasks are accomplished, users get a different perspective on a subset of the data. Changing the spatial layout of a node-link diagram by applying a layout algorithm is one example.
\begin{itemize}
\item Move selected nodes (mostly composite): This task is related to node-link diagrams where users want to change individual node positions interactively. Often, this can be accomplished with drag and drop interaction using the mouse cursor or a previously selected tool.
\item Adjust graph layout (always stepped): This task is performed when users want to apply a certain layout algorithm to a graph presented in a node-link diagram view. In the majority of systems, this can be achieved by selecting a layout algorithm from a list and pressing a button afterwards.
\end{itemize}

\paragraph{Encode Tasks:} The users intention when performing encode tasks is to alter the visual representation of the data including the visual appearance of each data item. Changing the color-mapping of graph edges in a matrix representation is one example.
\begin{itemize}
\item Change node size (mostly stepped): This task is accomplished when users want to alter the default range of node sizes. This task is supported with interactive sliders for example.
\item Change label size (mostly stepped): A common way to present node or edge identifiers is to use labels. Changing the label size can be necessary when they occlude each other, overlap nodes or edges, or when they are too small to read. This task is supported with drop-down menus where users can select predefined label sizes for example.
\item Change node/edge mapping  (mostly stepped): When changing the node or edge mapping, users aim to alter the visual representation of node/edge attributes, for example new attributes can be communicated that have not been visualized before. This can be commonly achieved through GUI dialogues.
\item Color node/edge independently from mapping (always stepped): This task is performed when users want to mark a node/edge in the visualization with a color different to the mapped color, e.g., to communicate a certain characteristic. Systems typically support this task with coloring-tools.
\end{itemize}

\paragraph{Abstract \& Elaborate Tasks:} By performing these tasks, users adjust the level of abstraction of a data representation. For example by expanding nodes of a clustered graph in order to see more details.
\begin{itemize}
\item Expand/Collapse node (always stepped): To expand a node means to drill down the view of the graph on a selected node of a clustered graph in order to view the associated subgraph. Collapsing a node is the reverse operation. These tasks are usually supported by selecting options in context menus.
\end{itemize}

\paragraph{Filter Tasks:} When filtering data, the users intention is to change the set of displayed data items based on specific conditions. Filtering is usually applied to nodes or edges when working with graphs.
\begin{itemize}
\item Apply node/edge filter (varying between all three): Change the set of displayed nodes or edges depending on specific conditions. Some systems provide dialogues where users are able to set filter conditions and provide sliders to apply the filters interactively.
\end{itemize}

\paragraph{Connect Tasks:} With connect tasks users highlight associations and relationships between data items that are already presented and show hidden information relevant to a specified item.
\begin{itemize}
\item Show/hide labels (always stepped): This task is performed when users want to show or hide labels in the visualization (e.g., node labels). This task is supported with a simple toggle button for example.
\item Show node/edge attributes (always stepped): Sometimes users need to view attribute data associated with a specific node or edge in a tabular form. These information are usually displayed in a separate view after selecting node or edge of interest.
\item Show metrics/statistics (always stepped): With this task users want to view metrics or statistics concerning the whole data set or a previously selected subset. Usually such information are provided in individual panels or views.
\end{itemize}

\begin{figure*}
\centering
\subfigure[Interactive Zooming]{
\label{subfig:zoom}  
\includegraphics[height=0.14\textheight]{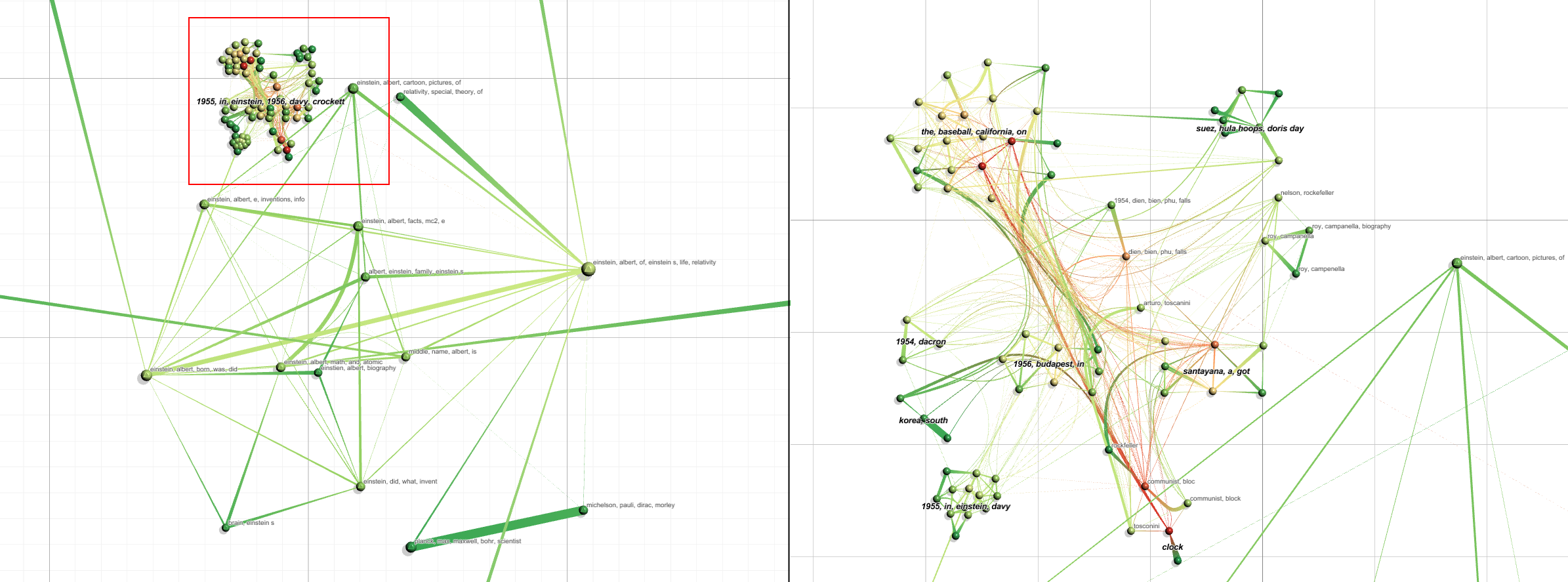}}
\subfigure[Expanding a node]{
\label{subfig:expand}  
\includegraphics[height=0.14\textheight]{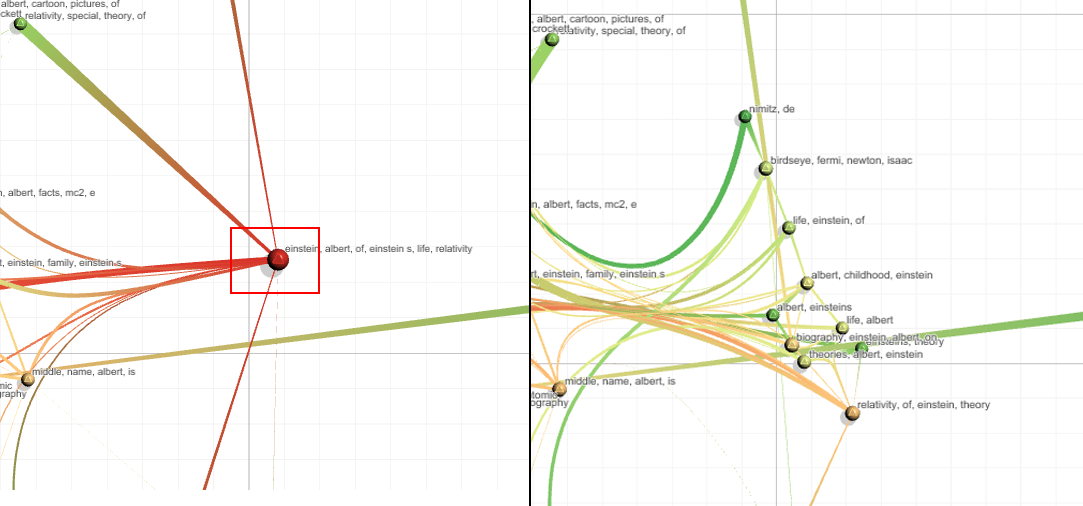}}
\caption{
Two interaction techniques for graph exploration tasks in CGV: (a) interactive zooming using the mouse cursor and wheel to specify the zoom center and zoom level, respectively (zoomed region marked with red rectangle), (b) Expanding a node (marked with red rectangle) of a clustered graph using a double mouse click.}
\label{fig:exploretasks}
\end{figure*}

To conclude, all categories from Yi et al.~\cite{Yi2007} are supported by graph visualization systems. Surprisingly, the selection of edges is not widely supported.
Concerning the interaction mode it becomes clear that mostly stepped interaction followed by composite interaction is used. The continuous interaction mode is used rarely. This is typical for systems developed for desktop environments. Figure \ref{fig:exploretasks} shows interaction techniques for zooming and expanding a node exemplarily.

\subsection*{Basic Tasks for Graph Editing}

\paragraph{Create Tasks:} By performing these tasks, users create empty documents for new data sets.
\begin{itemize}
\item Create empty document (always stepped): A task to create a file for a new data set. This task is usually supported with a GUI dialogue.
\end{itemize}

\paragraph{Insert Tasks:} When insert tasks are accomplished, new data items are added to the data set. Concerning graphs, inserting nodes and edges are examples.
\begin{itemize}
\item Insert node/edge (mostly composite): The task to insert a new node/edge. This task can often be accomplished by dragging a template shape from a shape palette to the visualization.
\item Insert copied node/edge/subgraph (always stepped): With this task a previously copied node/edge/subgraph shall be inserted. This is usually supported by right-clicking the visualization and selecting an option from a context menu.
\item Duplicate node/edge/subgraph (always stepped): With this task a node/edge/subgraph is copied and inserted again at once. After selecting the node/edge/subgraph, this can be accomplished by selecting an option from a menu for example.
\item Add node/edge attribute/label (always stepped): The task to add a node/edge attribute or label is also often supported with the use of GUI dialogues that can be invoked through a node specific context menu.
\item Add group to selected nodes/edge (always stepped): A task to group selected nodes/edges, for instance, to express a semantic relationship. One common interaction technique is to first select the nodes/edges and apply the grouping through a selection in a context menu afterwards.
\end{itemize}

\paragraph{Delete Tasks:} Delete tasks are performed in order to remove existing data items from the data set. Deleting individual nodes, edges or subgraphs are examples concerning graphs.
\begin{itemize}
\item Delete node(s)/edge(s)/subgraph (always stepped): With this task the user's intent is to remove selected objects from the data set. This can usually be done by selecting these objects first and pressing the delete button on the keyboard afterwards.
\item Remove group (always stepped): A task to remove a previously added group. This can usually be performed with right-clicking the group and selecting the dedicated option of a context menu.
\end{itemize}

\paragraph{Update Tasks:} By performing update tasks, users change characteristics of data items, for instance an attribute value of a graph node.
\begin{itemize}
\item Update node/edge attribute value (always stepped): A task for changing the value of a specific node or edge attribute. This can usually be performed with right-clicking the node/edge, selecting the dedicated options of a context menu and entering a new value with the keyboard.
\item Update node/edge label (always stepped): This task is performed similar to the task above, but for node or edge labels.
\end{itemize}

\paragraph{Navigate Tasks:} By performing these tasks, users navigate to different subsets of the data in order to edit them.
\begin{itemize}
\item Pan view (always continuous): This task is performed similar to the task in graph exploration.
\item Zoom view (mostly stepped): This task is performed similar to the task in graph exploration.
\end{itemize}

\begin{figure*}[t!]
\centering
\subfigure[Inserting an edge]{
\label{subfig:insert}  
\includegraphics[height=0.09\textheight]{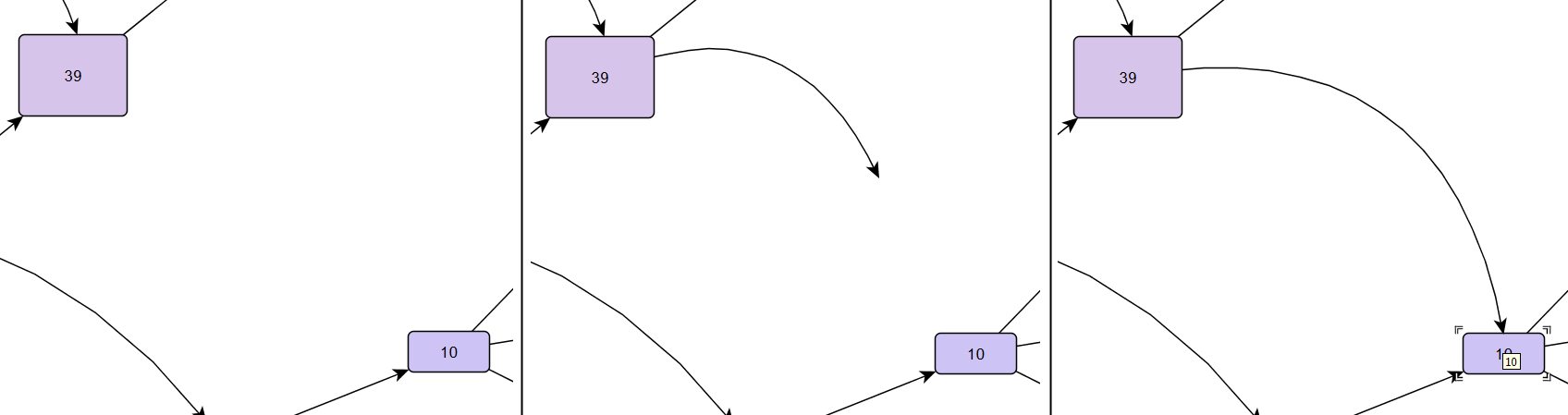}}
\subfigure[Deleting multiple nodes]{
\label{subfig:delete}  
\includegraphics[height=0.09\textheight]{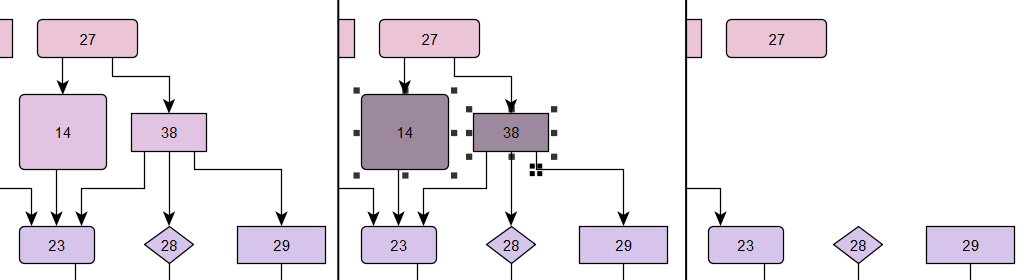}}
\caption{
Two interaction techniques for graph editing tasks in yEd: (a) Inserting an edge of an edge template palette with mouse drag and drop interaction, (b) deleting a set of nodes with the keyboard input \textsc{del} after selection with a rectangle frame. Individual steps of each interaction technique are illustrated from left to right.}
\label{fig:edittasks}
\end{figure*}

\paragraph{Select Tasks:} By performing selection tasks in this context, users mark data items of interest to edit them, e.g., nodes, edges, subgraphs, associated data attributes or sets of these.
\begin{itemize}
\item	Select node (mostly stepped) / Deselect node (always stepped): These tasks are performed similar to the task in graph exploration.
\item	Select edge (mostly stepped) / Deselect edge (always stepped): These tasks are performed similar to select node task but for edges.
\item	Select multiple nodes (varying between all three) / Deselect multiple node (always stepped): These tasks are performed similar to the task in graph exploration.
\item	Select multiple edges (varying between all three) / Deselect multiple edges (always stepped): These tasks are performed similar to select multiple nodes task but for multiple edges.

\end{itemize}

\paragraph{Miscellaneous Tasks:} Additional Tasks that do not fit in above categories.
\begin{itemize}
\item Copy node(s)/edge(s)/subgraph(s) (always stepped): A task to copy an object or a set of objects in order to insert them afterwards to the same or another data set. This can usually be achieved by selecting the objects first followed by the keyboard shortcut \textsc{ctrl + c} or a selection of a context menu option.
\item Cut node(s)/subgraph(s) (always stepped): A task to copy and remove objects at once. This is also mostly achieved with the use of context menus.
\item Change edge path (always composite): With this task users want to change the path of an edge. This is usually achieved by selecting an edge and dragging edge control points.
\end{itemize}

We identified basic tasks for graph editing in seven categories. Noteworthy, the navigation and selection tasks also belong to editing because pan, zoom and select/deselect are widely supported in graph editors as well. The reason for that is graph editors use node-link diagrams to display subsets of the graph and users need to navigate to different parts of the graph in order to view but also to edit them. Furthermore, users need to specify the objects to be affected by editing operations. The most common way to achieve this is with selection techniques. In contrast to systems for visual graph exploration, the selection of edges is widely supported in systems for graph editing. Another difference is, that temporary selections are usually not supported.
Concerning the interaction modes, we can identify a similar trend to the basic interaction tasks in visualization systems for graph exploration. Stepped interaction is used for the majority of tasks. Only very few systems utilize continuous or composite interaction modes. Because all the analyzed graph editors are desktop systems, this result is expected.
Figure \ref{fig:edittasks} illustrates interaction techniques for the insert edge and delete nodes tasks examplarily.

After identifying basic tasks for a system supporting graph exploration and editing, the next step is to determine what interactions can be used in order to accomplish these tasks. This is discussed in the following section.

\section{Interaction Vocabulary}
\label{vocabulary}

There is a wide variety of research in examining the possibilities to influence computer systems using modern human-computer interactions. 
These are mainly based on the concept of reality-based interaction~\cite{Jacob2008}, where the user's skills and capabilities of understanding basic physics, their body, their environment, and the social context are in focus.
In the following, we discuss different aspects and dimensions relevant to generate an extensive interaction vocabulary capable of distinguishing enough gestures to map onto the tasks presented in the previous section.

\subsection*{Interaction Modality}
Various interaction modalities have been explored in research, e.g.,~pen interaction, touch and multi-touch, tangible interaction~(actuated or passive tangibles), tangible views and spatial interaction, mid-air gestures, speech interaction, gaze-based interaction, brain-computer interfaces, multi-device interaction, proxemic interaction, and 3D interactions. 
While most of these can be used in context of interactive surfaces, we focus on the ones in actual contact to the surface: pen, touch and tangible interaction (see Figure~\ref{fig:modalities}).
However, it is of course possible to combine these contact-based interaction modalities with other contactless modalities, e.g.,~speech and touch~\cite{Tse2007}, or with each other, e.g., touch through tangibles~\cite{Buschel2014}, to improve interaction, its meaning and enrich the individual vocabulary. 
Additionally, individual modalities can specifically support a certain mental model of the user.
In a user-elicited study conducted by Frisch et al.~\cite{Frisch2009}, users specifically distinguished modalities for certain tasks, e.g.,~creation and editing tasks with pen, while manipulating elements through touch, while for other tasks pen and touch were used interchangeably, e.g.,~selecting a node. 

\subsection*{Single- / Multi-Device and Single- / Multi-User Gestures}
Within a single modality or through combining modalities, gestures in the vocabulary can be executed with different devices,~i.e.,~different fingers, hands, pens, or tangibles, and can thereby change the meaning of the gesture as long as they can be distinguished by the sensor system. 
A tap,~i.e.,~a short contact with the surface by a device, can be interpreted differently when made by a touch or pen, but also when performed with different devices within one modality, e.g., index finger or ring finger. 
Additionally, a concrete gesture can be made out of individual base gestures of single or multiple devices.
We could, for example, construct a gesture that is defined by a tap from a finger of one hand in conjunction with a tap from a finger from the other hand.
Taking this principle further, gestures can be specifically set out to be performed by a single specific user or multiple users in conjunction if user recognition and distinction is supported.
This may seem to make interaction unnecessarily complicated but can be useful for security measures or for global changes to a system in scenarios where users work in parallel and might otherwise disrupt another person's workflow. 

\subsection*{Concrete Gesture and Motion}
The individual gesture is very dependent on the selected interaction modality. 
General parameters to consider for distinguishing different gestures and the user's intentions are 
the presence of an object or user, the duration of its existence, position, motion, pressure, size, orientation, and sequence~\cite{Saffer2008}.
For continuous gestures motion can further be defined by the velocity of (parts of the) motion, the direction of movement, the path of movement or changes in direction during the motion. 
This section presents an overview of these parameters for the selected modalities.

\begin{figure}
	\centering
		\includegraphics[width=0.3\columnwidth]{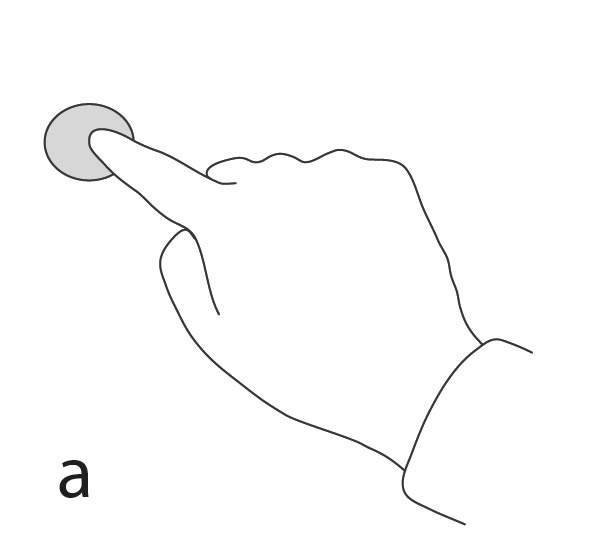} 
		\includegraphics[width=0.3\columnwidth]{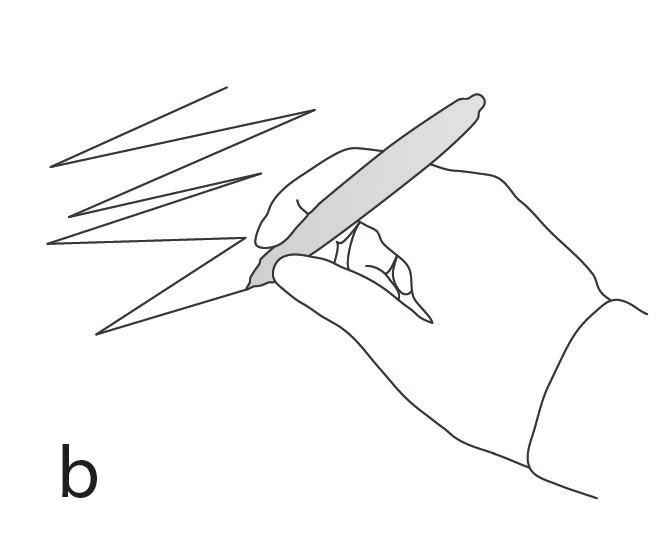} 
		\includegraphics[width=0.3\columnwidth]{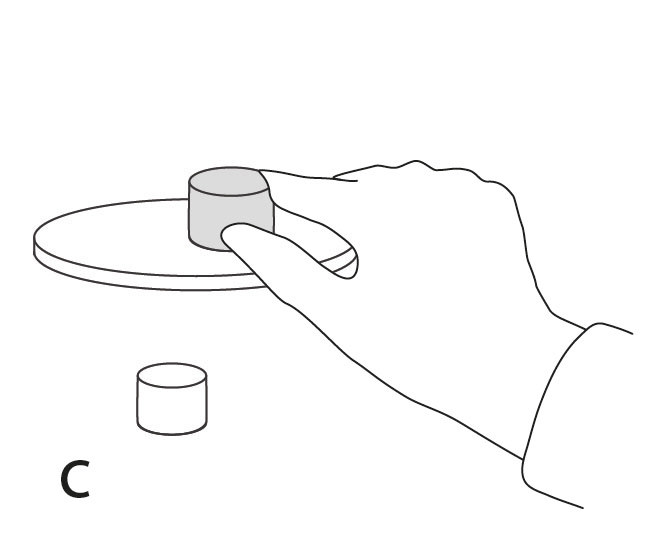}
	\caption{Contact-based interaction modalities for interactive surfaces: a)~Touch Interaction, b)~Pen Interaction, and c)~Tangible Interaction.}
	\label{fig:modalities}
\end{figure}

\paragraph{Degrees of freedom for touch and pen gestures}
\label{touch-dimensions}
In the following, we list a set of dimensions to consider when designing touch gestures. 
These are adapted from the research of Wobbrock et al.~\cite{Wobbrock2009} who analyzed user-elicited gestures.
We present these dimensions for touch and pen simultaneously since both present a single, direct pointer interaction.

\begin{itemize}
\item Continuity/Flow: $ discrete \leftrightarrow continuous $,\\
e.g., a short contact with the surface (tap) vs. a longer continuous movement on the surface (drag)

\item Duration/Velocity: $ short \leftrightarrow long~duration $,\\
e.g., a short contact (tap) vs. a long contact with the surface (hold) or alternatively, a short movement in one direction (flick) vs. a longer movement on the surface (drag)

\item Nature of motion: $ symbolic, physical \leftrightarrow metaphorical, abstract $,\\
e.g., dragging along a path vs. drawing a letter

\item Linearity of movement: $ straight \leftrightarrow changes~in~direction $,\\
e.g., one fluent movement vs. crossing something out multiple times (see Figure~\ref{fig:modalities}b)

\item Combination of base gestures
\vspace{-0.1cm}
\begin{itemize}
\item Number of touch points: $ single~touch \leftrightarrow multi~touch $,\\
e.g., one finger drag vs two finger drag
\item Number of hands: $ uni-manual \leftrightarrow bi-manual $,\\
e.g., drag with one finger vs hold with one hand and drag with the other
\item Relation of movement: $ parallel \leftrightarrow divergent~movement $,\\
e.g. two-finger drag vs. two finger moving away from each other (pinch)
\end{itemize}

\end{itemize}

\paragraph{Degrees of freedom for tangible gestures}
\label{tangible-dimensions}
Similar to touch and pen, the presence of tangibles can convey a meaning and thereby trigger a reaction of the system.
However, this is especially the case, as tangibles can stay in place for a longer time while new gestures are performed. 
More than for touch, tangible interaction distinguishes the type of device or actuator that was used to perform a gesture as different tangibles are often associated with different functions. 
Factors of these gestures are therefore not limited to motion with the tangible, but properties of the tangible itself. 
The dimensions to consider when creating a tangible vocabulary are:

\begin{itemize}
\item Form, size and flexibility: $ thin, bendable \leftrightarrow thick, rigid $,\\
e.g., foils, plates, tokens, blocks and compound forms~\cite{Buschel2014}
\item Materials: $ color, haptics,~and~transparency $,\\
e.g., tangibles made of wood, plastic, or acrylic glass
\item Role and function, e.g., $ function \leftrightarrow parameter \leftrightarrow data $,\\
e.g., tangibles can be used for invoking functions, as physical controls changing parameters, or as representatives or data containers~\cite{Buschel2014}
\item Interaction with a single tangible
\vspace{-0.1cm}
\begin{itemize}
\item Lifting or placing the tangible
\item Translation along the surface plane with different movements (see also the dimensions of motion for touch)
\item Rotation and orientation
\item Tilting
\item Flipping
\item Shaking
\end{itemize}

\item Combination of base gestures
\vspace{-0.1cm}
\begin{itemize}
\item Type of tangible: $ same~type \leftrightarrow different~type~tangibles $,\\
e.g., the small wooden tangible has a different meaning than the large wooded block
\item Relation of tangibles, $ decoupled \leftrightarrow coupled~tangibles $,\\
e.g., next to each other or overlapped and stacked (see Figure~\ref{fig:modalities}c)
\end{itemize}

\end{itemize}

These parameters describe the individual gestures in a vocabulary which can again be extended by combining different modalities to form a gesture. 
It is therefore unreasonable to list all possible gestures. 
However, it is essential to be aware of these parameters and resulting possibilities when designing a gesture set for specific tasks.

\subsection*{Object relations}
\begin{figure}[t]
	\centering
		\includegraphics[width=0.23\columnwidth]{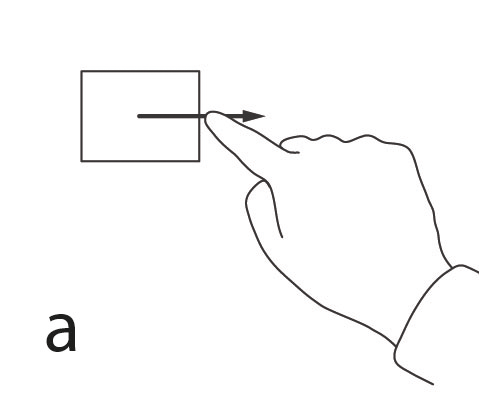}
		\includegraphics[width=0.23\columnwidth]{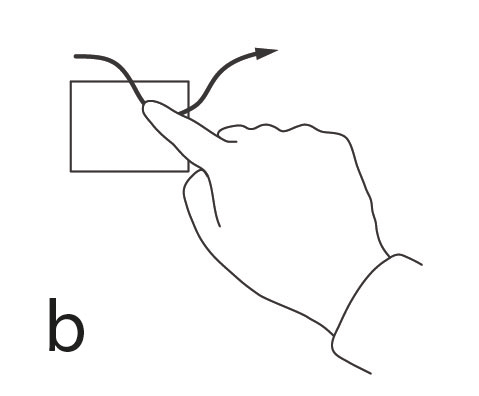}
		\includegraphics[width=0.23\columnwidth]{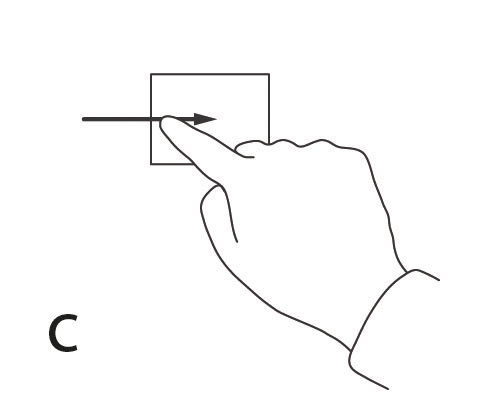}
		\includegraphics[width=0.23\columnwidth]{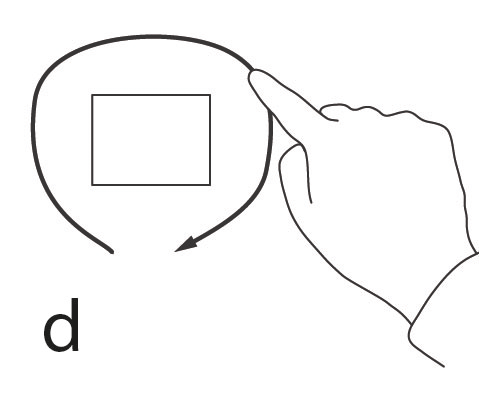}
	\caption{A gesture's meaning can be dependent on the relation to an object. Objects can be classified as part of the gesture when it a) started on it, b) crossed it, c) ended on it, and/or d) enclosed it.}
	\label{fig:object-relation}
\end{figure}

Gestures can be performed with respect to objects, to world features, or independent from the visualization~\cite{Wobbrock2009}.
While some of these features are of little to no importance for the intention the gesture conveys, often gestures are executed specifically in relation to an object.
Furthermore, not only fixed object but also different areas of the surface can be interpreted as objects with specific meaning, e.g., the area along the border of a view.
In terms of tangible interaction, a gesture can be performed in relation to a previously placed tangible. 
Then the tangible itself can function as a representative of an object.
Objects or areas that are part of the gestures can be categorized (see Figure~\ref{fig:object-relation}) as
\begin{itemize}
\item the object or area the gesture started on,
\item the object or area the gesture crossed,
\item the object or area the gesture ended on, and
\item the object or area that was enclosed by the gesture.
\end{itemize}

In summary, user actions have several degrees of freedom to be distinguishable from each other.
Interaction modality, device(s), concrete gesture and affected virtual object are only few examples as not all of them can be presented here. 
Having $n$ degrees of freedom $D_1, D_2, .., D_n$, the interaction vocabulary can be described as a set of $n$-tuples $I = \{i_1, i_2, ...,i_l\}$, where $i_j = (d_{j_1},d_{j_2},..,d_{j_n}), d_{j_k} \in D_p, 1\leq j \leq l, 1\leq p \leq n$.

So far we identified basic tasks for graph exploration and editing, discussed different aspects and dimensions to generate an interaction vocabulary and formally described it.
In order make users able to accomplish basic tasks interactively, a mapping from tasks to interactions has to be established. This is discussed next.
\section{Mapping Tasks to Interactions}
\label{matching}

After describing a set of basic interaction tasks $T = \{t_1, t_2, ...,t_k\}$ for graph exploration and editing, as well as an interaction vocabulary $I = \{i_1, i_2, ...,i_l\}$, the next step is to decide which interaction shall be used to accomplish which task. Hence, a mapping from tasks to interactions is needed.
Before providing a formal mapping approach, we describe pragmatic mapping examples for a small set of tasks.
 
\subsection*{Mapping Examples}

The following mapping examples use a small subset of basic interaction tasks from section \ref{tasks} and gestures generated from the parameters listed in section \ref{vocabulary}. 
We picked two exemplary mappings for each of the selected contact-based modalities: touch, pen, and tangible interaction. 
For use of touch and pen interaction in conjunction for graph creation and editing we refer to Frisch et al.~\cite{Frisch2009}. 

\subsubsection*{Interaction Modality - Touch}
In this example for a touch-based vocabulary mapped on basic graph interaction tasks, we present two gestures where one is specifically referencing a data object while the other is a global gesture influencing the whole view.
Because these basic tasks are either selective (focusing on a single node) or global (focusing on the view), the gestures are designed to either use single touch for the precise interaction and multi-touch for the interaction that has more extensive impact on the visualization.

\textbf{Select - Select Node $ \rightarrow $ Tap node}: \\
A single user shortly touches the surfaces in the location of the node,~i.e.,~start and end object equal the node to select.

\textbf{Explore - Center View $ \rightarrow $ Shake canvas}: \\
The user places multiple fingers on the surface, independent from any objects, then makes a back and forth movement in a way that all actuators frequently change their direction of movement in a short period of time.

\subsubsection*{Interaction Modality - Pen}
For the mapping with pen interaction, we selected an example showing the differences of memorized to symbolic, metaphorical gestures. 
While it would be possible to let the user draw a specific letter to indicate a commando, the drawing of symbolic features that indicate the resulting view is easier to recall from memory.

\textbf{Reconfigure - Apply graph layout $ \rightarrow $ Draw the result}: \\
In the right hand corner of the view, the user uses the pen to draw a simplified sketch of the intended layout, e.g., a ring for a circular layout or three nodes forming a tree for a tree layout.

\begin{figure}[h]
	\centering
		\includegraphics[width=0.32\columnwidth]{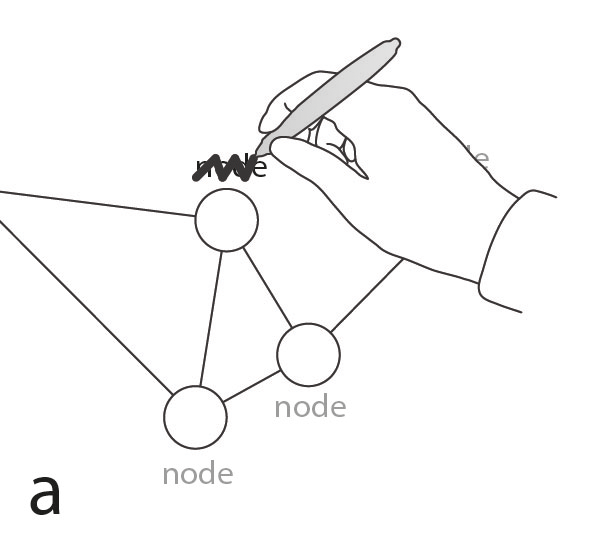}
		\includegraphics[width=0.32\columnwidth]{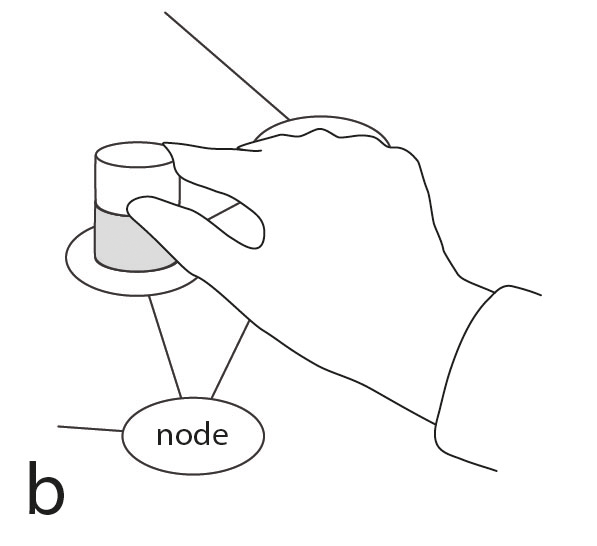}
		\includegraphics[width=0.32\columnwidth]{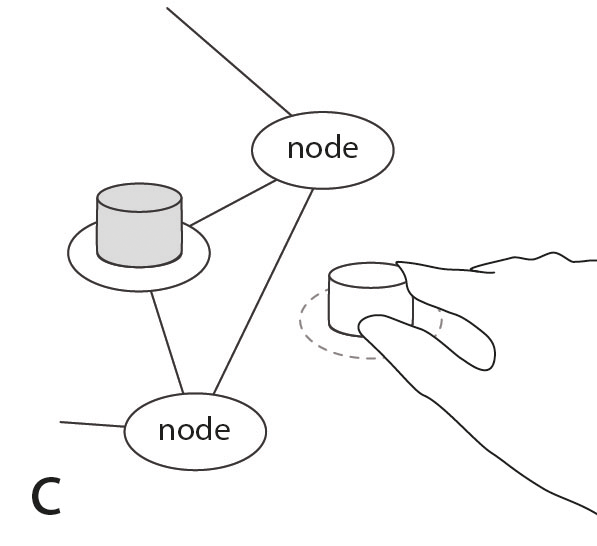}
	\caption{Exemplary mappings of tasks and gestures: a) Hiding label through crossing out a representative, b) to c) using tangibles to duplicate nodes similar to stamping.}
	\label{fig:mapping01}
\end{figure}

\textbf{Connect - Hide labels $ \rightarrow $ Cross out node and edge label}: \\
This gesture is composed from hiding the node and edge labels individually. For each, nodes and edges, one representative label is used where the user drags the pen in multiple consecutive motions with multiple changes in opposite directions of movement crossing the representative label (see Figure~\ref{fig:mapping01}a). By crossing out these labels, all labels will be hidden from view.

\subsubsection*{Interaction Modality - Tangible}
For the set of gestures forming the tangible interaction vocabulary, it is very convenient to assign specific categories to tangibles and thereby convey different purposes for each device. 
For this context, we assume an office setup where these tangibles can be stored.
In other cases, such as mobile applications, tangibles are rather inconvenient or have to be limited in their amount.
However, in any case the different devices have to be clearly distinguishable not only for the system, but the user. 
In our case, we assume a set of visibly different tangibles that are both small opaque tokens as well as larger transparent plates that can be placed on top of elements without occluding them.

\textbf{Insert - Duplicate node $ \rightarrow $ Stamp nodes}\\
One small transparent tangible for selection is placed on the node to be copied. Another tangible of this type is placed on top and then lifted up and placed on a free space in the canvas (see Figure~\ref{fig:mapping01}b-c). The second tangible functions as a stamp: A copy of the node appears in the location on which it is placed.

\textbf{Delete - Delete subgraph $  \rightarrow $ Select, then flip}\\
The user places a tangible for editing on the surface and selects the subgraph to be removed by moving the tangible, crossing the elements and connecting them to the tangible. When lifting up the tangible and flipping it, the elements are removed.\\

These examples indicate that finding a suitable mapping even for a small set of tasks is not trivial. The reason for that is the immense range of possibilities to consider. 
Having a larger set of tasks, finding a "good" mapping is even more difficult. In such cases we recommend to use a formal approach.
In the following, a formal approach for finding and evaluating a mapping is provided.

\subsection*{Formal Mapping Approach}
Mathematically, a mapping $m$ from tasks to interactions can be defined as $m: T \rightarrow I$. 
To avoid ambiguities, $m$ must be injective.
In order to set up a usable interaction alphabet, $m$ has to satisfy a set of certain criteria $C=\{c_1, c_2, ..., c_n\}$ often referred to as interaction guidelines or design rules \cite{Blackwell2001, Dix1997, Green1989, Kerren2013}.
Let $M$ be the set of possible mappings $F \rightarrow I$.
The quality of a mapping $m \in M$ concerning a specific criteria $c \in C$ can be expressed with a quality function $q: M \times C \rightarrow [0,1]$.
Hence, the overall quality of a mapping $m$ concerning $C$ can be expressed with the normalized sum $\hat{q} = \frac{\sum\nolimits_{i=1}^{n}q(m, c_i)}{n}$.
Considering varying criteria priorities in different application contexts, it makes sense to add weight factors to every summand so that the overall quality is expressed by $\hat{q} = \frac{\sum\nolimits_{i=1}^{n}\alpha_i \cdot q(m, c_i)}{n}$, with $\alpha_i \in [0,1]$.

Finding a "good" or "the best" mapping can now be interpreted as an optimization problem where the overall quality is maximized.

When applying this matching approach to a given set of tasks and interactions, a set of concrete matching criteria $C$ needs to be selected.
Common criteria in literature are:
\begin{itemize}
	\item Predictability: Interaction should always exhibit deterministic behavior for the user \cite{Kerren2013}.
	\item Consistency: Similar interactions should be used for similar functions \cite{Kerren2013}.
	\item Familiarity: Interaction should map as closely as possible to the real world or to known metaphors \cite{Kerren2013}.
	\item Generalizability: Interactions should be as specific to the context as necessary, but as basic as possible to be reusable in other contexts \cite{Kerren2013}. 
	\item Viscosity: Frequently used functions should map to interactions with lowest effort \cite{Blackwell2001, Green1989}.
	\item Recoverability: Users should easily undo interactions \cite{Dix1997}.
	\item Directness: Interaction should rather be directly applied to the affected virtual object than on separate control panels \cite{Shneiderman1983}.
	\item Continuity: Combination of basic interaction steps should be possible to form an interaction flow without discontinuities \cite{Elmqvist2011, Kerren2013}.
\end{itemize}


Moreover, the quality function $q$ has to be defined according to the semantics of the selected criteria.
An optimization algorithm can now be used to converge to a mapping with an optimal overall quality.\\


Even with this formal approach, setting up a mapping for the whole set of tasks described in section \ref{tasks} and the interaction modalities described in section \ref{vocabulary} still takes a considerable amount of time and effort.
Finding such a mapping is beyond the scope of this report and left for future work.

\section{Conclusion and Future Work}
In this technical report, we aim to support the development of a system for graph exploration and editing specifically designed for todays interactive surfaces.
With an extensive review of existing systems for either graph exploration or graph editing, we were able to define a list of basic interaction tasks that should be supported by a system for both graph exploration and graph editing.
In order to accomplish these tasks through user interactions, a mapping from tasks to interactions is necessary.
For this reason, we analyzed different interaction modalities according to their interaction vocabulary first and described additional degrees of freedom for interaction.
Since mapping tasks to interactions is difficult because of the immense range of possibilities and criteria to consider, we developed a general mapping approach.
By implementing this approach it becomes possible to set up a mapping with a good quality concerning these criteria.
With this work, we support future research in developing such a mapping.
Our entire approach is not limited to graph exploration and editing but generally applicable.

\section*{Acknowledgements}

This research has been supported by the German Research Foundation (DFG) in the context of the project GEMS -- graph exploration and manipulation on interactive surfaces.

\bibliographystyle{abbrv}

\bibliography{references}
\end{document}